\documentstyle[12pt]{article}
\makeatletter
\def\ps@myfrontpage{\let\@mkboth\@gobbletwo
 \def\@oddhead{\large\sl\vbox to0pt{\hbox{Department of Physics}
               \hbox{Kyoto University}\vss}\hfil
              \rm\vbox to0pt{\pubnumber\hbox{July 1994}\vss}}%
 \def\@oddfoot{}\def\@evenhead{}%
 \def\@evenfoot{}\def\sectionmark##1{}\def\subsectionmark##1{}}
\def\@maketitle{\newpage
\null
\vskip\temp
 \vskip 3em 
 \begin{center}
  {\large \@title \par}     
  \vskip 3em 
  {\large                        
   \lineskip .5em           
   \begin{tabular}[t]{c}\@author
   \end{tabular}\par}
\end{center}
\par
 \vskip 3em 
}
\makeatother
\def\lsim{\mathrel{\mathpalette\oversim<}}
\def\gsim{\mathrel{\mathpalette\oversim>}}
\def\oversim#1#2{\lower0.2ex\vbox{\baselineskip0pt\lineskip0pt
  \lineskiplimit0pt\ialign{$#1\hfil##\hfil$\crcr#2\crcr\sim\crcr}}}
\def\bx{{\bf x}}
\def\bk{{\bf k}}
\def\vgamma{\vec\gamma}
\def\np{n_p}
\def\bg{{\hat\gamma}}
\def\DTT{{\Delta T\over T}(\bx,\bg)}
\def\DTT1{{{\Delta T\over T}(\bx,\bg_1)}}
\def\DTT2{{{\Delta T\over T}(\bx,\bg_2)}}

\def\pubnumber{\hbox{KUNS 1279}
}
\newdimen\temp \temp=0pt
  \setbox1=\vbox{\pubnumber}\advance\temp by 1.5\ht1
\begin{document}

\title{\bf Skewness of CMB Anisotropies\\
             in an Inflationary\\
          Isocurvature Baryon Model}

\author{
Kazuhiro Yamamoto and Misao Sasaki\\
{\em Department of Physics, Kyoto University} \\
{\em Kyoto 606-01, Japan}}
\maketitle\thispagestyle{myfrontpage}
\vfill

\centerline{\bf Abstract}
\vspace{1pc}

We investigate the cosmic variance of the skewness of the cosmic
microwave background (CMB) anisotropies in an inflationary model which
leads to the baryon isocurvature scenario for the cosmic structure
formation. In this model, the baryon number fluctuations are given
by a sinusoidal function of a random Gaussian field.
We find that the skewness is very small in comparison with
that of the fluctuations which obey Gaussian statistics.

\vfill
\newpage

The existence of the nearly isotropic cosmic microwave background (CMB)
is a firm evidence of the hot big bang cosmology.
At the same time, it is believed that its anisotropies
carry valuable information about the early history
of our universe.
Recently, the COsmic Background Explorer (COBE) satellite team
reported the detection of temperature anisotropies in the CMB
on scales greater than $10^\circ$ (Smoot et al. 1992).
These fluctuations are particularly important because they directly reflect
the gravitational potential fluctuations at $z\sim1000$.
They are hence tightly related to the structure formation
of the universe, and their data are powerful clues to
know the origin of the cosmic structure.

The angular two-point correlation function is commonly used in a
statistical analysis of the temperature fluctuations.
However, its knowledge alone cannot distinguish the statistical
properties of the fluctuations. To do so, one needs to know at least
the next order correlation, i,e, the angular three-point
correlation.
In the case of the temperature fluctuations which originate
from the adiabatic curvature perturbations in the inflationary cosmology,
it is shown that its imprint on the angular three-point correlation
function is very small (Falk et al. 1993; Gangui et al. 1993),
and that the contribution from the cosmic variance will dominate over the
signal (Srednicki 1993; Scaramella 1991).

Recently the COBE team reported the results of the data analysis
of the three-point correlation function based on the DMR first-year
sky map (Hinshaw et al. 1993). In terms of a Monte Carlo analysis
with an assumed level of the instrumental noise,
they found an evidence of the
non-vanishing three-point correlation function in the data
which is consistent with the prediction of the standard adiabatic
cold dark matter (CDM) model based on the inflationary cosmology.
However, it is also true that the low signal-to-noise level in the data
prevented them to reproduce the three-point correlation function
in a clear form, and hence they could only place an upper limit on
the amplitude of it as a firm conclusion.
 According to them,
further several-year observations will produce clear data.

In this situation, it is worthwhile to consider the angular
three-point correlation in other possible scenarios of the
structure formation besides the standard CDM scenario.
Among them, a viable alternative is the Peebles isocurvature baryon
(PIB) scenario of the structure formation, in which the universe
consists of only baryonic matter and radiation but no
non-baryonic dark matter (Peebles 1987),
based on the observation $\Omega_0\sim0.1$ (Peebles 1986; Tyson 1988).

 Furthermore, the second year COBE-DMR data reported recently indicate
a power-law index of the primordial density fluctuations which is higher
than that of the Harrison-Zeldovich spectrum (Bennett et al. 1994).
In addition, several other recent CMB experiments on degree scales
have reported the detection of a high amplitude temperature
anisotropy (Gundersen et al. 1993; Hancock et al. 1994),
which may be regarded as an evidence in favor of the PIB scenario
(Hu \& Sugiyama 1994),
though not all of the results of degree scale experiments seem to
be consistent, which may be due to non-Gaussian statistics (Luo 1994).

However, the shortcoming of the PIB scenario, in the theoretical sense, is
that it assumes a very {\it ad hoc\/} spectrum of the primordial density
fluctuations to explain the structure formation and to satisfy
the observed isotropy of the CMB on large angular scales,
namely isocurvature baryon fluctuations with a steep spectral index.

Now, an interesting microscopic mechanism to produce the seemingly
{\it ad hoc\/} spectrum for the PIB scenario has been proposed
(Sasaki \& Yokoyama 1991; Yokoyama \& Suto 1991)
In this mechanism, the soft CP violation is induced by a spatially-varying
Majoron field, $A(\bx)$, associated with a heavy Majorana lepton field
which decays three quarks or three anti-quarks, and the
space-dependent net baryon number,
\begin{equation}
\label{Bsin}
  B(\bx)=B_* \sin\left({A(\bx)\over f}\right)
\end{equation}
is generated, where $f$ is the mass scale of the Majoron field
and $B_*$ is a constant determined by the coupling constants of
the model. In particular, Sasaki and Yokoyama (1991) found
that a power-law inflation model provides the most appropriate class of
power spectra for the PIB scenario.
In this case there appears a characteristic scale $k_c$,
which is determined by the model of inflation and the mass scale
$f$, and the power spectrum approaches the white noise as $k/k_c\rightarrow0$
and almost the scale invariant one as $k/k_c\rightarrow\infty$.
Thus it has a nice feature that the amplitude of
small scale fluctuations does not become too large on very small scales.
 Further, they have shown that there exists a natural particle
physics model that can provide an appropriate initial
condition for the PIB scenario, implemented in power-law inflation
with the power-law index around $\np\simeq 10\sim20$.
The only difference from the original PIB scenario is that
one naturally expect the presence of a cosmological constant that makes
the universe spatially flat in this inflationary isocurvature baryon model.
Detailed statistical properties of the baryon fluctuations in this model
has been studied by Yamamoto et al. (1992).
Considering the viability of the PIB scenario, it is important to
examine if this model for the PIB scenario has statistical properties
which are observationally testable.

In this letter, based on the results of Yamamoto et al. (1992),
we consider the skewness of the CMB temperature fluctuations in this scenario.
We denote the temperature fluctuation field by
${\Delta T\over T}(\bx,\vgamma_1)$, where $\bx$ specifies the position
of the observer, the unit vector $\vgamma_1$ points a given direction
from $\bx$. The temperature fluctuation can be evaluated using the
isocurvature perturbation theory (Kodama \& Sasaki 1986).
The baryon number fluctuations correspond to entropy perturbations
which give rise to gravitational potential perturbations after the
universe becomes matter-dominated. Then the temperature fluctuations
directly trace the primordial baryon number fluctuations on large
angular scales,
\begin{equation}
\label{DTB}
{{\Delta T\over T}(\bx,\vgamma_1)}=c B(\bx_1),
\end{equation}
where $c$ is a proportional constant (Sasaki \& Yokoyama 1991),
$\bx_1:=\bx+r_0\vgamma_1$,
$r_0=H^{-1}\int_0^1 dy/\sqrt{\Omega_0 y+(1-\Omega_0)y^4}
\sim2\Omega_0^{-0.4}H_0^{-1}$ ($\Omega_0\gsim0.1$),
and $H_0$ is the Hubble constant. Thus the statistics of
the temperature fluctuation can be understood through that of
$B(\bx)$.

Now we focus on the skewness of the CMB anisotropy,
\begin{equation}
\label{defskew}
  S:=\int {d\Omega_{\vgamma_1}\over4\pi}
  \biggl( {{\Delta T\over T}(\bx,\vgamma_1)} \biggr)^3.
\end{equation}
where $d\Omega_{\vgamma_1}$ denotes the integration over the
direction $\vgamma_1$. As is clear from Eqs.(\ref{Bsin}) and
(\ref{defskew}), we have $\Big<S\Big>=0$,
where $\Big< \ \ \Big>$ denotes the ensemble average on the
position of observer $\bx$. Then we consider the cosmic
variance of the skewness, which can be written as
\begin{equation}
\label{cvsa}
  \Big<S^2\Big>
  =\int\int{d\Omega_{\vgamma_1}\over4\pi}{d\Omega_{\vgamma_2}\over4\pi}
  \Bigg<\Biggl({{\Delta T\over T}(\bx,\vgamma_1)}\Biggr)^3
        \Biggl({{\Delta T\over T}(\bx,\vgamma_2)}\Biggr)^3\Bigg>.
\end{equation}
The ensemble average of the temperature fluctuations can be
replaced by the correlation of the baryon number fluctuations
by using Eq.(\ref{DTB}).
The random field $B(\bx)$ follows a peculiar statistic,
because it is a sinusoidal function of
the Gaussian random field $A(\bx)$. However,
we can evaluate the $2m$-point correlation function of $B(\bx)$
by using the following formula that is derived in by Yamamoto et al. (1992),
\begin{equation}
\label{cynssy}
  \Bigg<\prod_{j=1}^{2m} B(\bx_j)\Bigg>\simeq
  \biggl[{B_*^2\over4}e^{\np\beta}\biggr]^m
  \sum_\sigma {}' \exp\biggl[ \np\beta\sum_{j<i}\sigma_j\sigma_i
  \biggl({|\bx_i-\bx_j|^2\over \eta_f^2}\biggr)^{1/\np}\biggr],
\end{equation}
where $\beta:=H_f^2/8\pi^2f^2$, $H_f$ is the Hubble
parameter at the time when the inflation ends,
$\np$ is the power-law index of the power-law inflation,
$\sigma_i \ (i=1,2,\cdots, 2m)$ takes the value $\pm1$,
and $\sum_\sigma'$ means that the summation is taken all over the
combinations $(\sigma_1,\sigma_2,\cdots,\sigma_{2m})$ with
$\sum_{j=1}^{2m} \sigma_j=0$.
After straightforward but tedious calculations, we find,
\begin{eqnarray}
\label{BBBBBB}
  \Bigg<B(\bx_1)^3~B(\bx_2)^3\Bigg>=
  {1\over4}\left({B_*^2\over2}\right)^{-6}\Bigg<B(\bx_1)B(\bx_2)\Bigg>^9
  +{9\over4}\left({B_*^2\over2}\right)^{2}\Bigg<B(\bx_1)B(\bx_2)\Bigg>.
\end{eqnarray}
Note the appearance of the 9th power of the two-point correlation in
the first term. This term arises because there are $3\times3$ independent
combinations between $\bx_i$ ($i=1,2,3$) and $\bx_j$ ($j=4,5,6$) before
we take the coincidence limits.
The two-point correlation of $B(\bx)$ is given by (Sasaki \& Yokoyama 1991)
\begin{equation}
\label{BB}
  \Bigg<B(\bx_1)B(\bx_2)\Bigg>=\int {d^3\bk\over(2\pi)^3}
  e^{i\bk(\bx_1-\bx_2)}P_B(k),
\end{equation}
where
\begin{equation}
\label{PB}
  P_B(k)=2\pi B_*^2 e^{\np\beta} {1\over k^3}
  \int_0^\infty ds s \sin s
  \exp\left[-\Biggl({s\over k/k_c}\Biggr)^{2/\np}\right].
\end{equation}
The ensemble average of the angular two-point correlation function
is given by $C(\alpha):=\Big<{\Delta T\over T}(\bx,\vgamma_1)
{\Delta T\over T}(\bx,\vgamma_2)\Big>=c^2\Big<B(\bx_1)B(\bx_2)\Big>$,
where $\cos\alpha=\vgamma_1\cdot\vgamma_2$.
This is written in terms of the multipole moments as
\begin{equation}
\label{Calpha}
  C(\alpha)={1\over4\pi}\sum_{l} (2l+1)
  \Biggl(c^2{2\over\pi} \int_0^\infty dk ~k^2 j_l(kr_0)^2 P_B(k)\Biggr)
  W_l^2 P_l(\cos\alpha).
\end{equation}
Here the window function $W_l=exp(-l(l+1)/2\sigma^2)$,
$\sigma=17.8$, is inserted in order to compare it with the result
of COBE (Smoot et al. 1992).

Then, from Eqs.(\ref{cvsa}),(\ref{BBBBBB}) and
(\ref{Calpha}), we obtain
\begin{equation}
  \Big<S^2\Big>= {1\over8} \Bigl({B_*^2\over2}c^2\Bigr)^{-6}
  \int_{-1}^{1}d (\cos\alpha) ~C(\alpha)^9,
\end{equation}
where we have used the fact that the monopole term
(and also the dipole term) in the temperature
fluctuations should be subtracted from the data.
The variance of the skewness in the Gaussian statistics takes a
similar form (Srednicki 1993), but differs from the present
case in the powers of $C(\alpha)$.

 For comparison with other papers
we normalize the variance of skewness by $C(\alpha=0)^3$, and
consider the normalized root mean square skewness, {\it i.e.,}
$\hat S_{r.m.s.}:=\big<S^2\big>^{1/2}/C(0)^{3/2}$.
We can show that this can be written as
\begin{equation}
\label{rmss}
  \hat S_{r.m.s.}=
  \biggl({\sqrt{2}\over\pi}e^{\np\beta}\biggr)^3 {\cal A},
\end{equation}
where
\begin{eqnarray}
\label{defA}
  &&{\cal A}:=\Biggl[{\int_{-1}^1d(\cos\alpha)
  \Bigl\{\sum_{l}(2l+1)d_lW_l^2P_l(\cos\alpha)\Bigr\}^{9}\over
  \Bigl\{\sum_{l}(2l+1)d_lW_l^2\Bigr\}^{3} } \Biggr]^{1/2},
\\
\label{defdl}
  &&d_l:=\int_0^\infty{dk\over k} j_l(kr_0)^2
  \int_0^\infty ds s \sin s
  \exp\left[-\Biggl({s\over k/k_c}\Biggr)^{2/\np}\right].
\end{eqnarray}
 For the present scenario to be successful, we assume
$\np\beta< 1$ (Sasaki \& Yokoyama 1991). Hence, the coefficient of
${\cal A}$ in the Eq.(\ref{rmss}) is of the order of unity.
Therefore we focus on the numerical factor ${\cal A}$.
If the two parameters, {\it i.e.}, $\np$ (the exponent of the
power-law inflation) and $x_c:=r_0k_c$ (the ratio of the present
horizon size $r_0$ to the characteristic scale $1/k_c$) are specified,
we can calculate it numerically.
The Table~1 shows the results of numerical integration of ${\cal A}$
for various values of $x_c$ in the case $\np=10$\, and
the Table~2 does in the case $\np=20$.

Once $\np$ and $x_c$ are fixed, we can also obtain the angular
two-point correlation function $C(\alpha)$ from Eq.(\ref{Calpha}).
We numerically fitted this angular two-point correlation function
$C(\alpha)$ normalized by $C(0)$ to that due to the density fluctuation
with the power-law spectrum
$\Big<\Bigl(\Delta\rho /\rho\Bigr)_k^2\Big>\propto
k^{n_{\rm eff}}$, and calculated the best fitted value of $n_{\rm eff}$.
Thus $n_{\rm eff}$ is the effective spectral index of the fluctuations
on the COBE scale; $n_{\rm eff}\gsim2$ for $x_c\gg1$ while
$n_{\rm eff}\sim1$ for $x_c\sim1$.

Let us roughly evaluate an upper limit on ${\cal A}$.
In the case $x_c\sim1$, for which the spectrum is almost
scale invariant, we have $d_l\simeq\pi/2l(l+1)\np$. Then we find
${\cal A}\lsim O\bigl(\np^{-3}\bigr)$. However, this case is not
appropriate for the PIB scenario.
When $x_c\gg1$, which is the case of our interest,
the value of ${\cal A}$ is very small as shown in the Tables~1 and 2.
Thus the cosmic variance of the skewness turns out to be very small
in comparison with the one obtained for the Gaussian statistics with
the Harrison-Zeldovich spectrum, $\hat S_{r.m.s.}\simeq0.18$ (Srednicki 1993).
This suggests that the cosmic variance of the angular three-point
correlation function is also very small.
We suspect that the higher order correlation is always
suppressed in this model.

If we accept the data analysis of the COBE team (Hinshaw et al. 1993),
 {\it i.e.},
the detection of the nonvanishing three-point correlation as
they claim, it is difficult to explain it within the present model of
the PIB scenario.
However, either the actual amplitude of the skewness or the form of
the three-point correlation function has not been obtained
due to the high noise level. According to them,
the noise level of the three-point correlation function diminishes
in proportion to $(\hbox{\it time})^{-3/2}$ as the data accumulate.
It is necessary for us to wait for a few more years before a definite
conclusion may be drawn.

\vskip 10mm

\begin{center}
\underline{\bf Acknowledgement}
\end{center}
K.Y. is grateful to Professor H. Sato for his encouragement.
This work was supported in part by Monbusho Grant-in-Aid for
Scientific Research No.05640342, and the Sumitomo Foundation.

\newpage

\begin{center}
\underline{\large \bf References}
\end{center}
\begin{itemize}

\item[] Bennett,~C.~L., et al. 1994, Preprint, astro-ph/9401012,
submitted to ApJ.
\item[] Falk,~T., Rangarajan,~R. \& Srednicki,~M. 1994, ApJ, 403, L1.
\item[] Gangui,~A., Lucchinm,~F., Matarrese,~S. \& Mollerach,~S. 1993,
Preprint \\ SISSA REF.193/93/A, DFPD 93/A/80, astro-ph/9312033.
\item[] Gundersen, J.~O. et al. 1993, ApJ, 413, L1.
\item[] Hancock, S. et al. 1994, Nature, 367, 333.
\item[] Hinshaw,~G. et al. 1993, Preprint, astro-ph/9311030.
\item[] Hu, W. \& Sugiyama, N. Preprint CfPA-TH-94-16, UTAP-178,
astro-ph/9403031.
\item[] Kodama,~H. \& Sasaki,~M. 1986, Int. J. Mod. Phys. A, 1, 265.
\item[] Luo,~X. 1994, Preprint CfPA-Th-22, astro-ph/9404069.
\item[] Peebles,~P.~J.~E. 1986, Nature, 322, 27.
\item[] Peebles,~P.~J.~E. 1987, Nature, 327, 210.
\item[] Sachs,~R.~K. \& Wolfe,~A.~M. 1967, ApJ, 147, 73.
\item[] Sasaki,~M. \& Yokoyama,~J. 1991, Phys. Rev. D, 44, 970.
\item[] Scaramella,~R. \& Vittorio,~N. 1991, ApJ, 375, 439.
\item[] Smoot,~G.~F. et al. 1992, ApJ, 396, L1.
\item[] Srednicki,~M. 1993, ApJ, 416, L1.
\item[] Tyson,~J.~A. 1988, Astron. J., 96, 1.
\item[] Yamamoto,~K., Nagasawa,~M., Sasaki,~M., Suzuki,~H. \&
Yokoyama,~J. 1992, Phys. Rev. D, 46, 4206.
\item[] Yokoyama,~J. \& Suto,~Y. 1991, ApJ, 379, 427.

\end{itemize}

\newpage
\renewcommand{\arraystretch}{1.2}
\centerline{Table~1}
\vspace{3mm}
\centerline{Skewness and effective spectral index for $\np=10$.}

\begin{center}
\begin{tabular}{c@{\hspace{3pc}}c@{\hspace{3pc}}c}
\hline\hline
 { \ \ $x_c$ \ \ } & {\ \ ${\cal A}$ \ \ } & {\ \ $n_{\rm eff}$ \ \ } \\ \hline
$10^6$ & $3.2\times 10^{-13}$ & 2.6  \\
$10^5$ & $3.0\times 10^{-9}$ & 2.1 \\
$10^4$ & $1.4\times 10^{-6}$ & 1.7 \\
$10^3$ & $5.5\times 10^{-5}$ & 1.4 \\
$10^2$ & $3.6\times 10^{-4}$ & 1.17 \\
$10$ & $7.3\times 10^{-4}$ & 1.04 \\
\hline
\end{tabular}
\end{center}

\vspace{8pc}

\centerline{Table~2}
\vspace{3mm}
\centerline{Skewness and effective spectral index for $\np=20$.}

\begin{center}
\begin{tabular}{c@{\hspace{3pc}}c@{\hspace{3pc}}c}
\hline\hline
 { \ \ $x_c$  \ \ } & {\ \ ${\cal A}$ \ \ } & {\ \ $n_{\rm eff}$ \ \ } \\
\hline
$10^{11}$ & $2.8\times 10^{-14}$ & 1.9 \\
$10^9$ & $6.5\times 10^{-10}$ & 1.6 \\
$10^7$ & $3.7\times 10^{-7}$ & 1.33 \\
$10^5$ & $1.0\times 10^{-5}$ & 1.17 \\
$10^3$ & $5.6\times 10^{-5}$ & 1.07 \\
$10$ & $9.6\times 10^{-5}$ & 1.01 \\
\hline
\end{tabular}
\end{center}

\end{document}